# Synchronized Attachment and the Darwinian
# Evolution of Coronaviruses CoV-1 and CoV-2

J. C. Phillips

Dept. of Physics and Astronomy, Rutgers University, Piscataway, N. J., 08854

## Abstract

CoV2019 has evolved to be much more dangerous than CoV2003. Experiments suggest that structural rearrangements dramatically enhance CoV2019 activity. We identify a new first stage of infection which precedes structural rearrangements by using biomolecular evolutionary theory to identify sequence differences enhancing viral attachment rates. We find a small cluster of mutations which show that CoV-2 has a new feature that promotes much stronger viral attachment and enhances contagiousness. The extremely dangerous dynamics of human coronavirus infection is a dramatic example of evolutionary approach of self-organized networks to criticality. It may favor a very successful vaccine. The identified mutations can be used to test the present theory experimentally.

The current coronavirus belongs to a virus family that also causes common colds. CoV-1 and CoV-2 are much more dangerous. Worse still, many numbers now show that CoV-2 (2019) spikes have evolved to be much more dangerous than CoV-1(2003) spikes. Much of this "nearly perfect" 2019 viral action [1] can be explained in terms of two well-studied cleavage-reassembly structural differences of coronal spikes [2]. The spikes contain $> 1200$ amino acids, and BLAST shows that ~ 300 of these are mutated. Can only a few of these 300 mutations, far from the two cleavage sites, also be contributing to the extremely strong viral interaction of CoV-2? Maybe, but how do we identify these possibly critical distal and so far unobserved sites?

Here we will show that the exceptional viral strength of CoV-2 [1] can be used to find such sites. Our search confirms the importance of the two known cleavage sites [2]. It selects from the nearly 300 remaining mutated sites one new small mutated cluster, which can also contribute to the very strong viral activity of CoV-2 and precede cleavage. Our search is based on the concept of self-organized criticality, which goes back to van der Waals (1873) and has been discussed extensively by physicists for decades, but with only a few applications to proteins [3].

The present application uses a technique called hydropathic scaling, which has successfully explained small sequence differences in the evolutions of many protein families The thermodynamic principles of self-organized criticality are universal, and have successfully produced strong signals from the evolution of many proteins [4-11]. The shape of a protein depends strongly on its ins and outs relative to its center. The limits of these ins and outs are associated with hydrophobic and hydrophilic extrema. The shape of the protein changes during phase transitions. The success of hydropathic scaling often depends on the difference between first- and second-order phase transitions. Protein unfolding from water to air is a first-order transition, well-described by the most popular hydropathic scale (KD) [12]. Second-order transitions involve much smaller structural (long-range or allosteric) transitions, such as may occur in viral cellular attachment and penetration.

Second-order phase transitions are characterized in principle by fractal (non-integral) exponents, which are difficult to measure. The discovery of 20 universal amino-acid specific fractals in the solvent accessible surface areas (SASA) of > 5000 protein segments [13] is by far the most important application of phase transition theory in the 21[st] century and the latter part of the 20[th] century. The existence of these fractals proves that proteins function near critical points of second-order phase transitions, with human proteins often closest to their critical points [5] with level sets [10]. As the reader has guessed, the fractal hydropathic SASA scale $\Psi(aa)$ (MZ) is the tool that enables us here to find the new distal site cluster which probably contributes to the very strong viral activity of CoV-2.

The standard tool for comparing two or more proteins is BLAST, which compares different protein sequences site-by-site. It is not enough to profile $\Psi(aa)$ site by site for two proteins, as this produces hundreds of oscillations, with no clear differences between the two proteins. Instead we plot $\Psi(aa,W)$, where W is a rectangular box of length $W = 35$ over which $\Psi(aa)$ is averaged . This W value is chosen to maximize the hydropathic shape differences between CoV-1 and CoV-2, as measured by their variance ratio, which is dominated by extrema [11]. Smoothing the profile over W enhances resolution by reducing the number of extrema to the level which corresponds to critical domain motion differences. Choosing the best value of W to enhance resolution of evolutionary differences is similar to adjusting the focal plane of a microscope.

Our main result is shown in Fig. 1, which displays the 400-800 central region of the spikes. This region is part of the N-terminal domain (S1) responsible for cellular attachment [14]. Fig. 1 shows the $\Psi(aa,W)$ hydropathic profiles of for CoV-1 and CoV-2, using the $\Psi(aa)$ values of MZ [4]. While the two $\Psi(aa,35)$ profiles are similar, there are important differences in their extrema. The maxima are most hydrophobic and are located in the globular interior, while the minima are most hydrophilic and are located on the globular surface. The two cleavage sites S1/S2 and $S_2$· [2] of CoV-1 have moved lower (hydrophilically, further outside) in CoV-2 (Fig. 1), consistent with the very accurate MZ scale. When a cleavage segment is further outside, there is more space for cleavage and reassembly, which will occur more rapidly. The insertion PRRA in CoV-2 was identified [2] with BLAST ($W = 1$) as unique to CoV-2, but with BLAST alone one cannot show that this change has made CoV-2 more dangerous.

A characteristic feature of second order phase transitions is that they break what physicists often call a "hidden" symmetry, here the key symmetry of extrema [15]. Proteins are typically composed of domains of ~ 100 amino acids, and these domains can rotate from the resting state to the functional state, and then return reversibly to the resting state. This phase transition requires a balance between stability and flexibility [16,17]. To maximize the reversibility and extend protein life one can synchronize this domain motion by leveling the pivotal hydrophobic (inside) extrema forming level sets [10,11,18-21]. Viruses must act rapidly before being destroyed by antibodies, and they could do this through synchronized motion differently, by leveling their hydrophilic (outside) extrema. As shown in Fig. 1, such a leveling of minima 1-3 occurs in CoV-2, while it is absent from CoV-1. The change in minimum 2 is especially striking: it is caused by a cluster of four critical mutations from CoV-1 to CoV-2. These are (CoV-1 site numbering from Uniprot P59594): 546Gln to Leu; 556 and 561Ser to Ala; and 568Ser to Leu. The differences associated with each of these mutations are hydropathically large (~50-100 in the MZ scale [4]; all 20 amino acids span a range from most hydrophilic to most hydrophobic of 170).

How important is the fractal MZ scale? In Fig. 2 the MZ panoramic profiles of CoV-1 and CoV-2 are compared across the entire 1255 amino acid spike. We see immediately a strong hydrophobic peak above 1200, which is responsible for anchoring the spike to the central cushion; it is almost unchanged from CoV-1 to CoV-2. The central hydrophilic level set, absent from CoV-1 and present in CoV-2, is our main result (Fig. 1). Similar plots with the first order KD scale, shown in Figs. 3 and 4, do not exhibit synchronization by a hydrophilic level set (tilted arrow). One can see a shorter and more nearly level hydrophobic set, but only in CoV-1 and not in CoV-2.

There is an excellent review of the principles of self-organized criticality in living matter [3]. It mentions examples of synchronization (also used here for the hydrodynamics of domain motion of molecular level sets [15,18]). Another example is melting phase transitions in DNA strands. These can be compared to relaxation of homogeneous glass alloys, whose composition has been adjusted to bring the glass network to a critical point. There the glass (commercial Gorilla glass) is strong and flexible – not brittle [21]. Another broad and deep review includes fractals and their many implications for quantifying protein dynamic criticality [22]. In the genomic age abstract tools have many applications to analyzing the evolution of protein functions.

Some readers may be interested in the connections between hydropathic scaling theory of proteins and the more general synchronization of complex networks. The analogy is closest for networks that are scale-free because they are near a second-order phase transition. Social networks contain many such examples, usually described by one or two fractals [23]. Generally it is found that synchronizability is robust against random removal of nodes, but it is fragile to specific removal of the most highly connected nodes. In proteins the most highly connected nodes are hydrophobic extrema (deeply interior), where the number of van der Waals contacts is largest. In coronavirus spikes, it is the hydrophilic extrema that are most exposed to water and most likely to attach to protein targets. Thus the symmetry of extrema is broken between hydrophobic (proteins) and hydrophilic (CoV-2). One can also regard Darwinian selection as a kind of percolation [24]. Further extensions to multiphase hydrodynamic level sets require a special theory [15].

Our primary focus here has been on enhanced attachment caused by the synchronization of CoV-2 hydrophilic minima, but the mutations of CoV-1 are of some interest (P59594). Only a fragment (318-510) of CoV-1 accounts for its binding, and smaller fragments (327-510) and 318-490) do not bind [25]. The successful fragment is centered on the hydrophilic minimum 1 in Fig. 1, and it spans the region on both sides up to the two nearest hydrophobic maxima in Fig. 2. It was found that hydroneutralizing (alanining, Ala) six extremely hydrophobic (Cys) or hydrophilic (Asp, Glu) sites caused five of them to lose their binding: 452, 454, 463, 467, 474. These grouped sites are all close to the hydrophilic minimum 1 in Fig. 1 (MZ scale) and Fig. 3 (KD scale). Fragmental binding is a feature common to both scales, and thus not sensitive to whether the phase transition is first- or second-order. Instead, by synchronizing minima 2 and 3 with minimum 1, CoV-2 is able to greatly increase its attachment to ACE2 (angiotensin-converting enzyme 2) and probably other receptors as well; synchronization is specifically sensitive to allosteric interactions and is definitely a second-order effect.

In conclusion, the answer to the question posed in [1] is that, compared to CoV-1, CoV-2 has moved closer to its functional critical point [3]. CoV-2 has an added synchronized feature in its central attachment region that contributes to its extreme contagiousness in an asymptomatic phase [26]. The attachment of the S1 region precedes cleavage [2], and it is likely that both mechanisms enhance the infectiveness of CoV-2. An early estimate of the basic contagious reproductive number (R0) of Coronavirus 2019 was 2.2-2.7, but this was later revised to 5.7 [27]. By comparison, the numbers for H1N1 flu (1918, 1.8; 2009, 1.5) were much smaller [28] The uncontrolled R0 for CoV-1 was 3.6, and this dropped to 0.7 after the rapid implementation of control measures [29].

The new feature explains the unexpected yet often asymptomatic behavior of early stages of CoV-2 infection, when the new feature is acting alone. It should be possible to desynchronize this added attachment feature and reduce R0 by attaching a small molecule in the region 530-575. In any case, the "hidden" symmetry of the CoV-2 sequence should be of interest to both biologists and physicists. The effects of the four critical mutations on CoV contagiousness could be tested on mice. If the predictions of their importance should prove to be correct, then the general principle of Darwinian evolution of proteins would be confirmed.

The level hydrophilic extrema of the spikes of CoV-2 may be unique among viruses. These level extrema bring CoV-2 contagion very close to a critical point. There very small differences in DNA can cause large fluctuations in infection levels, not only between individuals, but also even in neighboring countries [30]. The quasi-linear spike morphology, with its large surface/volume ratio, is ideal for strong protein-water interactions, explaining [1]. The Oxford vaccine is based on the CoV-2 spike protein hydrophobically anchored to an adenovirus vaccine vector [31-33]. It seems likely that this vaccine will be more effective for CoV-2 than for CoV-1 because the synchronized attachment mechanism is stronger for CoV-2 and thus more easily disrupted by induced antibodies. Thus vaccines based on the entire spike are expected to be more effective than vaccines based on only part of the spike [34].

The most dangerous type of flu virus is $H_3N_2$, and it does not exhibit level sets [35]. Flu viruses mutate significantly annually, while CoV-2 has shown only a few mutations [36]. The spike stability against mutations is consistent with maximized attachment. There are many examples of scale-free networks. Readers unfamiliar with the general properties of scale-free networks, such as (often only 1 or 2) fractals, preferential attachment, and synchronization, may find early reviews interesting [37,38]. The sequences used here for CoV-1 and CoV-2 are the ones used in [2], that is, AAP13441.1 and YP_009724390.1. Many other CoV sequences are aligned (W = 1) in their Supp. Table [2]. Similar W = 1 sequence alignments have been reported by others [39,40], but the genetic origin of CoV-2 remains mysterious [1,40].

Historically it has long been the case that the amount of information that could be obtained from hydropathic scales was limited, because the fractals describing second-order phase transitions were not known. Now that we are in the genomic age, with a very large sequence data base available for many proteins and many species, the discovery of these 20 fractals [5,13] opens a new biophysics field of accurate thermodynamic analysis of small but medically important evolutionary differences.

The methods used here on CoV-1 and CoV-2 do not require elaborate calculation to obtain new results. The hydropathic scaling methods are easily implemented on a spread sheet, but are still not routine, and vary between proteins. The modular nature of protein domains responsible for protein assembly is well known [41,42]. Here we maximized small and otherwise mysterious evolutionary differences by using hydropathic scaling to optimize W at 35. This enables us to focus on domain-scale motion responsible for stronger attachment in CoV-2. Evidence for large-scale motion has been found in the evolution of many proteins [4-7,9], including molecular motors [11], while it is not accessible to traditional W = 1 BLAST-based alignment methods [2, 43-45]. From detailed genomic studies it has been suggested that "a more comprehensive theory of molecular evolution must be sought" [46]. Readers wishing to explore these novel scaling methods can obtain copies of sample EXCEL spread sheets from the author.

Many physicists have long suspected that proteins must function near a critical point very near thermodynamic equilibrium [3]. However, to implement this idea in practice with protein sequences, one needs to compress the existing vast array of three-dimensional structural data into a one-dimensional form. If you are one in a million, namely among the small community of co-authors of the 500+ references of [3], what comes next may not surprise you. As discussed above, this near-miracle compression has been achieved in the 21[st] century by two Brazilian physicists [13], located far from the main stream. [13] described the discovery of 20 (not just one) universal amino-acid specific fractals in the solvent accessible surface areas (SASA) of proteins. This amazing 21[st] century discovery was achievable only for proteins. It has made possible the quantitative analysis of Darwinian evolution of many different proteins [4-11], which brings us to the present application.

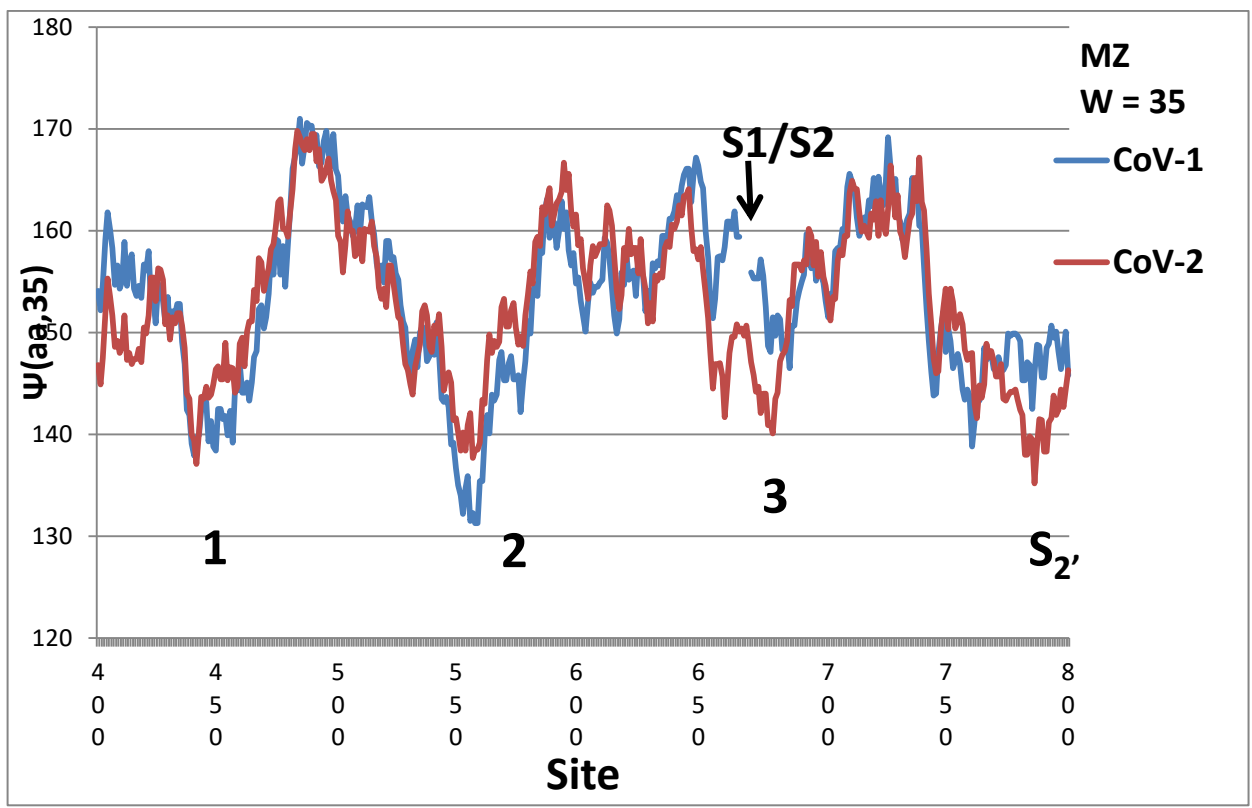

Fig. 1. The hydropathic profiles of CoV-1 and CoV-2 reveal a hidden symmetry when plotted using the MZ scale. The three hydrophilic minima of CoV-2, labeled 1-3, are nearly equal at 140, whereas the similar minima of CoV-1 range from 131 to 147. Note especially the very deep minimum of CoV-1 at 559. The new sequence PRRA in CoV-2, inserted at 681 in CoV-1 (the S1/S2 cleavage interface) [2], has an average MZ hydropathicity 108.5. This lowers $\Psi(aa,35)$ to 140.9 at the 3 minimum, aligning it with ~140 minima 1 and 2. The three minima span ~ 250 amino acids sites, which makes their water-driven synchronization for CoV-2, but not CoV-1, outside the range of most simulation or modeling methods.

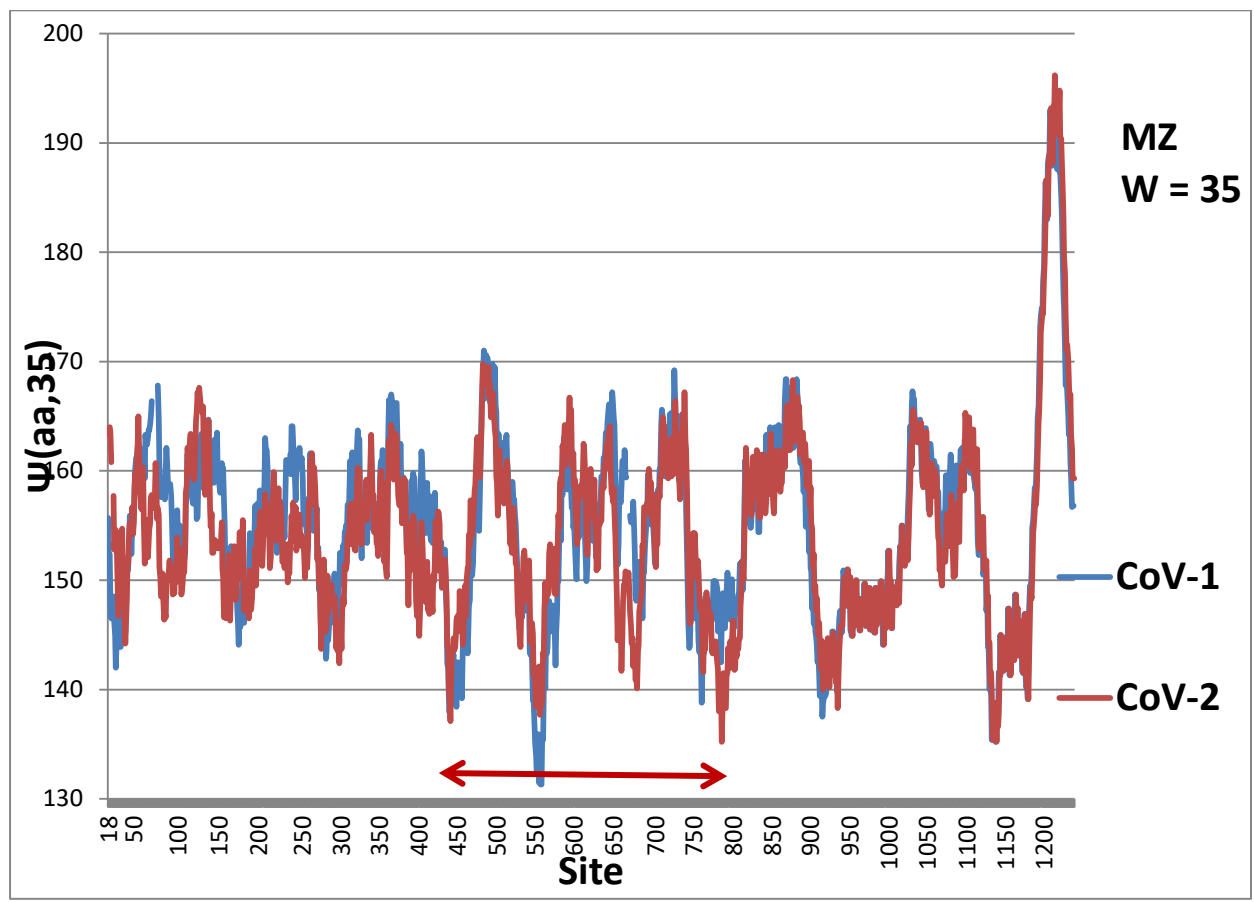

Fig. 2. With the MZ scale panoramic spike profiles of CoV-1 and CoV-2 reveal a set of hydrophilic level extrema in CoV-2, but not in CoV-1. The new level set was identified with a cluster of four single mutations in 546-568.

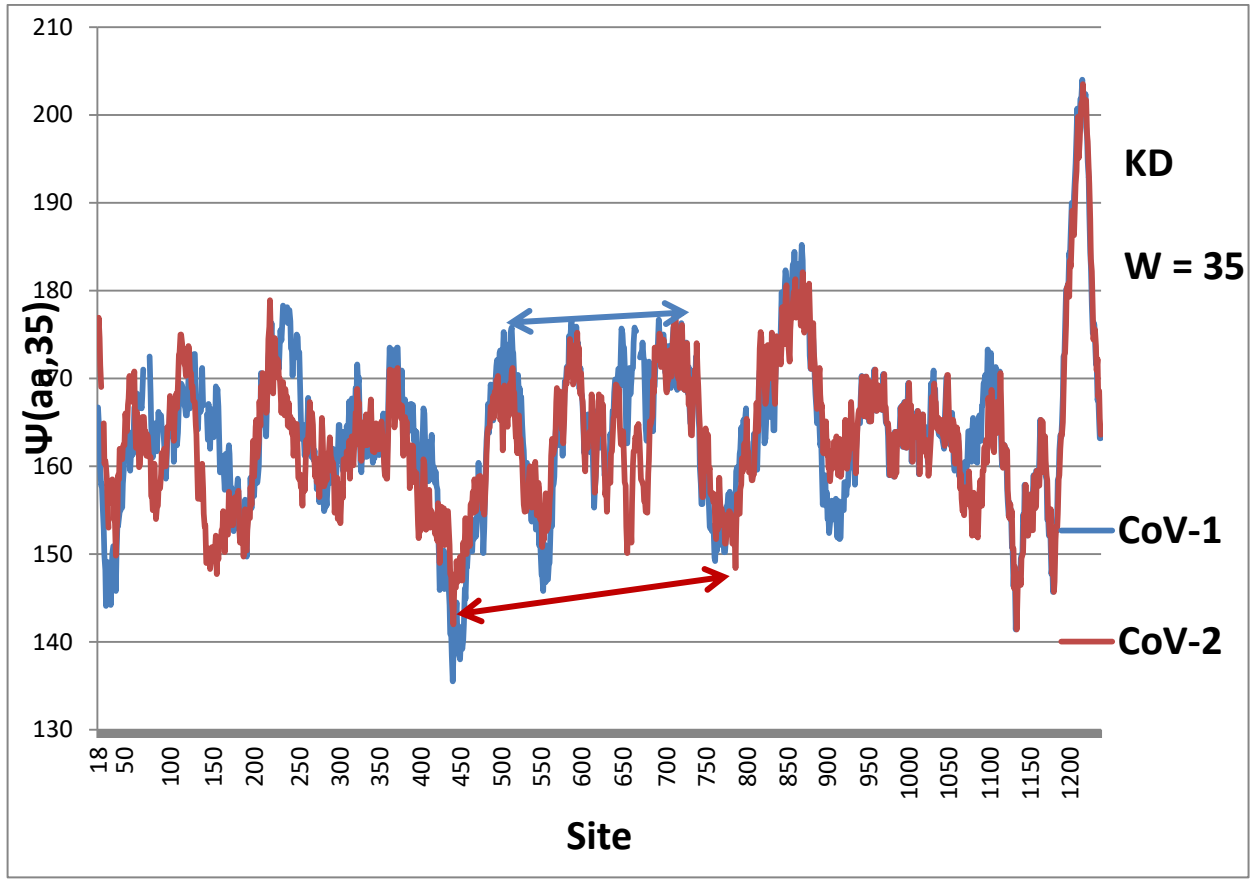

Fig. 3. Even on the panoramic picture, it is clear that with the KD scale the three minima that were aligned in CoV-2 are not level. This is shown in more detail in Fig. 4.

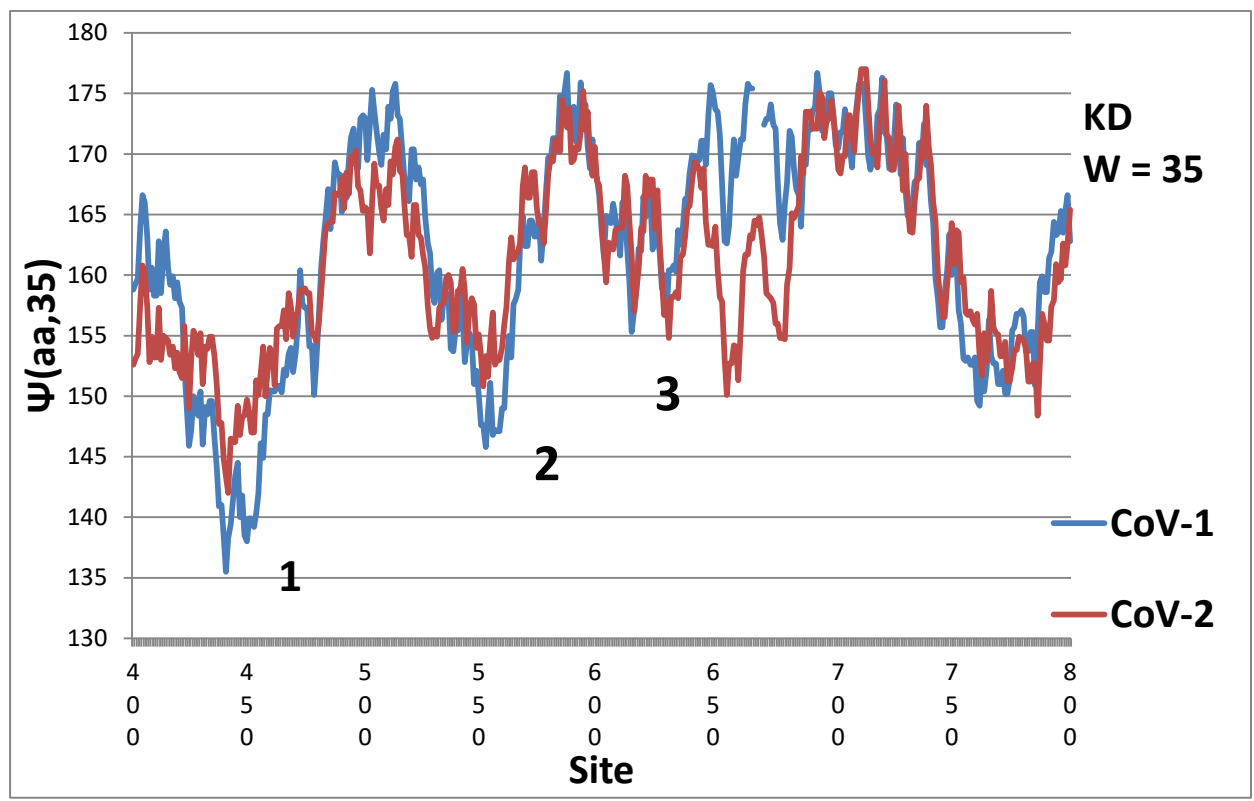

Fig. 4. The enlarged central region 400-800 of the hydropathic profile using the KD scale is similar to what was shown in Fig. 1 using the MZ profile, but there is an important difference. With the MZ scale all three minima 1-3 are aligned in CoV-2, but here only the 2 and 3 are aligned near 150. The 1 minimum has improved from 135 in Cov-1 to 142 in CoV-2, but this is not alignment, compared to minima 1-3 with the MZ scale (Fig. 1).